# Surface functionalization enhanced magnetism in $SnO_2$ nanoparticles and its correlation to photoluminescence properties

Venkataramana Bonu,[a] Arindam Das,[*a] Manas Sardar,[*b] Sandip Dhara,[a] and Ashok Kumar Tyagi[a]

High value of magnetic moment 0.08 emu/g at room temperature for $SnO_2$ nanoparticles (NPs) was observed. Surface functionalization with octadecyltrichlorosilane (OTS) enhanced the saturation magnetic moment of NPs to an anomalously high value of 0.187 emu/g by altering the electronic configuration on NPs surface. Surface functionalization also suppressed photoluminescence (PL) peaks arising from oxygen defects around 2 eV and caused an increase in the intensities of two peaks near violet region (2.6 - 3 eV). PL studies under uniform external magnetic field enriched understanding of the role of OTS. Both OTS and external magnetic field significantly modulated the luminescence spectra, by altering the surface electronic structure of NPs. Extra spins on the surface of $SnO_2$ NPs created by the surface functionalization process and their influence on resultant magnetic moment and luminescence properties are discussed in details.

## 1. Introduction

$SnO_2$ is a diamagnetic wide band gap (3.62 eV) *n*-type semiconductor. Irrespective of its wide direct band gap and large exciton-binding energy of 130 meV,[1] its dipole forbidden band-edge quantum states do not allow band to band transitions.[2] Consequently defect energy levels guide the optical transitions. Controlled modulation of defects, however, is need to make $SnO_2$ as a practical material for magnetic and optical applications. Annealing of amorphous $SnO_2$ thin films in optimized conditions showed sharp UV light emission which was used for diode application in the *n*-$SnO_2$/*p*-GaN heterostructure.[3]

Surface functionalization influences the electronic configuration of metal oxides and metals at considerable rate.[4-7] Electrical and optoelectronic performances of one dimensional ZnO nanostructures were improved by surface functionalization with organic molecules.[6,7] When Au nanoparticles (NPs) were coated with organic molecules (thiols), were found to be ferromagnetic even at room temperature.[8,9] Since then a large number of works on the organic material coated Au, Ag revealed that the ferromagnetism was intrinsic. The 'S' atom of thiol formed covalent bonds with Au atoms on the surface, which led to localization of *d*-hole around Au atoms, with well defined moment.[8,9] Surface functionalization with thiol, amine, tryoctylphosphine oxide (TOPO) molecules caused 10 nm diamagnetic ZnO NPs to gain magenic moment of 0.949, 0.158, 0.001 $\mu_B$ per NP, respectively.[10] Extensive experimental works on tuning of magnetic moment by coating on metal oxide NPs are reviewed by, Crespo *et al*.[9] More recently it is found that the induced moment due to functionalisation by organic molecules on various metal oxide semiconductors points to oscillated moment with surfactant concentration. The induced moment on an individual 4 nm $SnO_2$ oxide nanoparticle oscillates between 0.007 to 0.201 $\mu_B$/NP with TOPO and 0.007 to 0.0226 $\mu_B$/NP with poly (acryl acid) PAA functionalization, as increased amount of charges transfer from the surfactant molecules to the NPs. The magnetisation is also reported to saturate at very small fields of 2 KOe.[11]

Octadecyltrichlorosilane (OTS) is one of the widely used silanes as self-assembled monolayers, especially in organic semiconductors.[12,13] Here we report very high saturation magnetic moment in $SnO_2$ NPs (0.08 emu/g) for pristine 4 and 25 nm NPs. Moment per particle for 4 and 25 nm NPs is 2 and 492 $\mu_B$/NP, respectively. A simple one step surface functionalization with OTS renders further improvement in the saturation magnetic moment to 0.187 emu/g (4.7 $\mu_B$/NP) in 4 nm and 0.143 emu/g (879 $\mu_B$/NP) in 25 nm NPs. This change in moment is very large, and means that individual NPs carry very high moments when compared with reported values.[9] The mechanism of moment formation is clearly different from earlier known mechanisms. We have observed, large rearrangements in the spectral weights of the photoluminescence (PL) intensity of $SnO_2$ NPs due to surface functionalisation. Similar spectral rearrangemnts in the PL intensity occurs in pure $SnO_2$ NPs under a uniform external magnetic field (H = 0.1 T). We present a local quantum chemical modeling of bonding of OTS molecules to different sites on the surface allowing a localised spin ½ close to the binding site. This allows for building up of large moments in individual NPs. The identical role for both OTS, and external magnetic field, in modulating luminescence properties of $SnO_2$ NPs, validates our proposed model of moment formation and correlate it with ions to defect energy levels.

## 2. Experimental section

### 2.1 Materials synthesis

Stannic Chloride (SnCl$_4$, Spectro Chemicals) was mixed with distilled water (MilliQ, 18 MΩ) and was subsequently allowed to react with 0.05 mol/L ammonia solution (NH$_4$OH, Merck) for the synthesis of SnO$_2$ NPs. As-prepared NPs were annealed in air atmosphere for 1hr at 300 and 800 °C to get crystalline NPs which were dispersed in cyclohexane to promote surface functionalization with OTS (C$_{18}$H$_{37}$SiCl$_3$, Aldrich). Detailed procedures of synthesis and surface functionalization were reported elsewhere.[12]

### 2.2 Characterization techniques

Structural analysis was carried out by high resolution transmission electron microscopy (HRTEM; Libra 200 Zeiss). Raman spectroscopy was used to confirm the atomic bonding between OTS and SnO$_2$ NPs. Micro-Raman spectroscopy (InVia; Renishaw) using 514.5 nm excitation of Ar$^+$ laser is carried out with 1800 gr.mm$^{-1}$ grating and thermoelectric cooled CCD detector in the backscattering configuration. X-ray photoelectron spectroscopy (XPS; M/s SPECS GmbH, Germany) was employed to study the stoichiometry of the surface of NPs from the quantitative analysis of both Sn and O. The PL spectra were acquired using 325 nm line from He-Cd laser as the excitation source at RT. Temperature dependent measurements were performed in adiabatic stage (Linkam). Two permanent bar magnets were used to create uniform magnetic field. Magnetic properties of the samples were studied by the vibrating sample magnetometer (VSM-Mini cryo).

## 3. Results and discussion

### 3.1 Morphological and structural studies

Mean diameter of SnO$_2$ NPs annealed at 300 °C is observed to be ~ 4 nm from the TEM image and size distribution plot (Figs. 1(a) and b)). Fig. 1(c) shows the HRTEM image of crystalline (110) plane with $d$-spacing of 3.39 Å corresponding to rutile SnO$_2$ (JCPDS #41-1445) in the inset (zoomed). The ring like selected area electron diffraction (SAED) pattern (Fig. 1(d)) is indexed as (110), (101) and (211) planes corresponding to rutile SnO$_2$ phase with crystallites having all possible orientations. NPs annealed at 800 °C are found to have 25 nm as mean diameter. Detailed characterization of 25 nm size particles were reported in our previous study.[14] Functionalization of SnO$_2$ NPs with OTS offering superhydrophobic coating was also studied in one of our earlier reports.[12] Raman spectroscopy was employed for studying the atomic bonding between OTS and SnO$_2$ NPs (supplementary information in Fig. S1). The Raman spectra confirmed the surface functionalization of SnO$_2$ NPs by the OTS molecules exhibiting characteristic vibrational modes and related features as discussed in the supplementary information. OTS molecules are linked to SnO$_2$ NPs by forming covalent bonds following a self-assembly process. A plausible reaction mechanism between OTS and SnO$_2$ surface is shown in Fig. 2

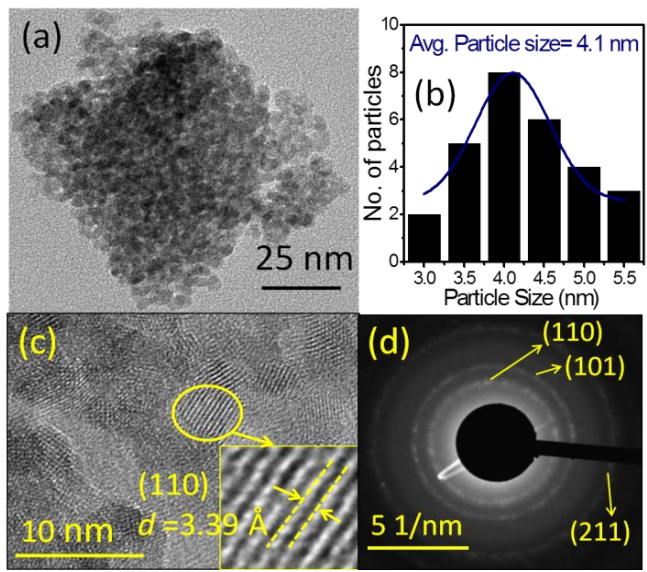

Fig. 1 Morphology and structural studies of SnO$_2$ NPs (a) Low magnification TEM micrograph (b) Particle size distribution showing peak at ~ 4 nm (c) HRTEM image with inset showing crystalline lattice plane of (110) with $d$ = 3.39 Å (d) SAED pattern showing rings corresponding to different crystalline planes.

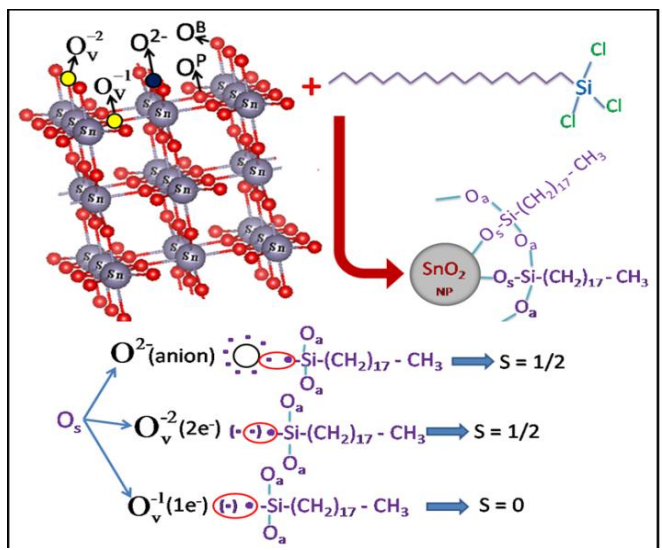

Fig. 2 Schematic image of surface functionalization process. Here, O$^P$: Planar oxygen; O$^B$: Bridging oxygen; O$_a$: Adsorbed atmospheric oxygen; O$_s$: Positions (O$^{2-}$, O$_v^{-2}$ and O$_v^{-1}$) on the SnO$_2$ nanoparticle surface where OTS are attached.

### 3.2 Magnetic properties

Room temperature (RT) magnetic hysteresis curves of the functionalized and pristine SnO$_2$ NPs were recorded using VSM (Fig. 3). Inset in Fig. 3 shows the zoomed region around 0.0 T. The RT saturation magnetization is about 0.08 emu/g at a field H = 0.5 T for both the size NPs. Surface stochiometry anlysis by XPS has revealed the 'O' to 'Sn' ratio in 4 and 25 nm as 1.7 and 1.55, respectively (Fig. S2). This indicates 'O' deficiency in 25 nm NPs surface is higher than the 4 nm NPs. PL measurements further supports this observation. Oxygen vacancy related PL peak intensity around 2 eV is higher in 25 nm NPs than 4 nm NPs (Fig. 4). Higher defects density in 25 nm NPs compensates lesser surface area in 25 nm NPs. As a result we see similar amount of moment per gram of both 4 and 25 nm NPs. This high defects density in 25 nm NPs is due to the desorption of 'O' from the surface of NPs while annealing as-prepared material at high temperature 800 $^{o}$C. Whereas 4 nm NPs were obtained by annealing the as prepared material at 300 $^{o}$C only. Magnetic moments were also observed in un-doped wide-band gap metal oxides like, $HfO_2$, $TiO_2$, $In_2O_3$, $Al_2O_3$, $CeO_2$ which are basically diamagnetic in the bulk form.[15] There is a general consensus now, that the magnetism in these oxides occurs from lattice defects like 'O' vacancies.[9] There are suggestions,[16] that moment in pristine $SnO_2$ NPs may come from Sn vacancies too.

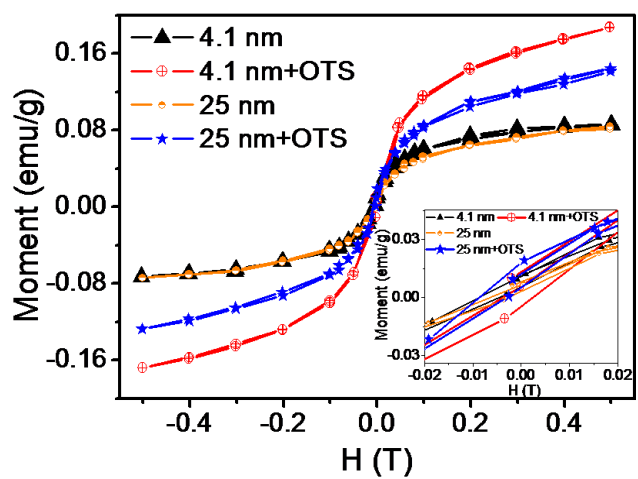

Fig. 3 Magnetic hysteresis curves of functionalized and pristine SnO2 NPs. Inset showing zoomed region around 0 T.

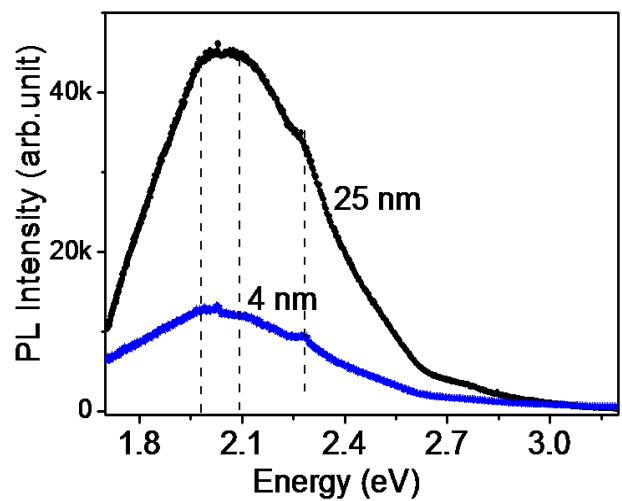

Fig. 4 PL spectra of pristine 4 and 25 nm NPs.

After functionalization of the particles, the RT saturation magnetization value increases to 0.18 emu/g at H = 0.5 T, i.e. the net magnetization doubles (Fig. 3). Bonding scheme[12] (Fig. 2) shows that a 'Si' atom forms 4 single bonds, one with a 'C' atom, two with oxygen atoms those are not part of the $SnO_2$ particle (adsorbed atmospheric oxygen atom, $O_a$) and only one with either an 'O' atom or 'O' vacancy (denoted as $O_s$ in Fig. 2) on the surface of the particle. Oxygen, in an ionic insulator like $SnO_2$, is in $O^{-2}$ state which is a filled shell. The formation of a single bond with a 'Si' atom leaves one extra unpaired electron on the 'O' atom. If the Si atom forms a singlet bond with one of the two electrons in a neutral oxygen vacancy ($O_v^{-2}$) site on the surface. Here also one electron is left with an extra spin $S = ½$ near the oxygen vacancy site. If the oxygen vacancy ($O_v^{-1}$) site is charged, on the other hand, containing only one electron, then the bonding with 'Si' actually leads to disappearance of spin $S = ½$. In our bonding scheme, each OTS molecule injects a local spin ½ on the surface of the particles, and each individual nanoparticle aquires large moments. Natural interaction between such injected spins due to localised electrons is antiferromagnetic superexchange interaction. Inset of Fig. 3 shows that there is an increase in coercive field with functionalisation. This seems natural because with increase in the number of spin ½ moments, the mutual interaction between the injected moments on the surface can lead to the increase in spin-spin correlation length which approaches to the particle size. In that limit, each nanoparticle will behave like a supermoment and induce hysteretic behaviour because of average anisotropy energy of each supermoment. Magnetic moment of 4.1 nm + OTS material is higher than the moment of 25 nm + OTS. Surface area of 4.1 nm NPs, in comparison with 25 nm NPs, is higher. This helps more OTS molecules to bind to 4.1 nm NPs than 25 nm NPs. Effect of this induced magnetisation on its defect related optical property like PL was also studied and is discussed in next section.

### 3.3 Photoluminescence properties and effect of magnetic field

PL studies performed on both the pristine and the functionalized $SnO_2$ NPs for defect origin of magnetization in the materials. Effect of magnetic field on the luminescence properties of these materials investigated to unravel the possible origin of magnetic moments. All PL studies were performed on 25 nm NPs which showed a band gap of (3.6 eV) in order to excite all the defect bands with He-Cd laser (325 nm) of 3.81 eV excitation.[14] Fig. 5 shows the PL spectra of 25 nm NPs for (a) pristine NPs, (b) OTS functionalized and (c) pristine $SnO_2$ NPs under uniform external magnetic field. A broad luminescence peak around 2 eV was observed for all three cases. Similar luminescence behavior for $SnO_2$ nanobelts, nanoribbons, nanowires was also reported.[17-21] Peaks at 2.77 eV and 2.97 eV were also present in pristine material with faint intensity (Inset of Fig. 5). A broad peak in the visible region of the spectra around 2 eV was suppressed and peaks in the violet region at 2.77 eV and 2.97 eV intensities were slightly enhanced in the presence of uniform magnetic field. Similar effect was observed pronouncedly in OTS treated NPs. The zoomed spectra for pristine sample in the absence and in the presence of magnetic field (inset of Fig. 5) display two clear peaks at ~ 2.77 and 2.97 eV. It is obvious from Fig. 5 and Fig. S4 (supplementary information) that there is no new peaks due to OTS treatment. OTS molecules only modulated intensities of the existing peaks in $SnO_2$. In order to investigate the different transition energies, all the curves were fitted using Gaussian and shown in Fig. 6 (a-d). Fig. 6(d) shows the gaussian fit of faint intensity peaks at 2.77 and 2.97 eV in the presence and the absence of magnetic field. It is convenient to discuss features in PL spectra for energies E < 2.6 eV and E > 2.6 eV. The PL spectra with E < 2.6 eV, can be easily deconvoluted, and it has essentially 5 broad lines with peaks at 1.84; 1.97; 2.12; 2.28; 2.44 eV (Fig. 6). The peak positions and fractional area of each peak are given in Table 1. Arrows in Table 1. indicates increase or decrease of fractional area as compared to pristine sample. The peak positions are not found to change in the functionalized NPs. Fractions of peak area for pristine $SnO_2$ to be, 11%, 28.3%, 30.1%, 20.6 % and 9.1% for electronic transitions with emitted photon energies centering, $E$ at 1.84, 1.97, 2.12, 2.28, 2.44 eV, respectively. After functionalization of these NPs (see Fig. 5), two incidents happen; (1) there is an overall reduction of the intensity in first four channels (see discussions later) but more importantly (2) the fractions corresponding to five peaks discussed above changes to 4.8%, 16.6%, 24.1%, 23.3% and 32.1%. The emission peaks at 1.97 and 2.12 eV are decreased and emission peaks at 2.28 and 2.44 eV channels are increased, after functionalization (Table 1).

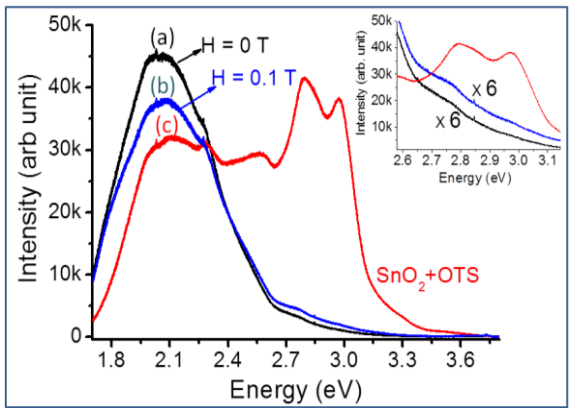

Fig. 5 PL of $SnO_2$ NPs (25 nm) at (a) H = 0 t (b) H = 0.1 T (c) functionalized with OTS

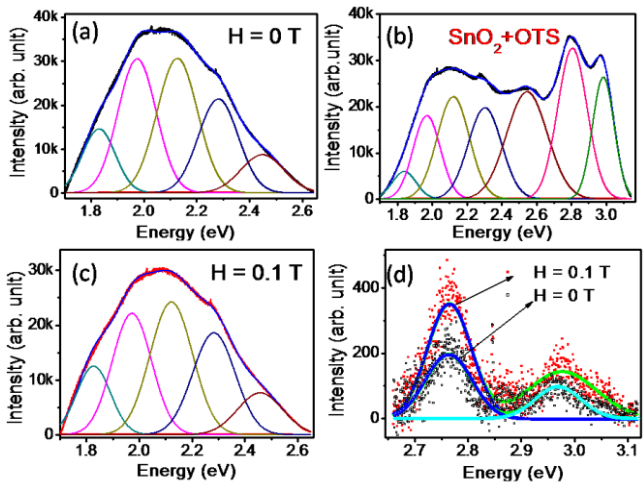

Fig. 6 Gaussian fitting of PL spectra of 25 nm $SnO_2$ NPs (a) Pristine NPs at H = 0 T (b) OTS treated NPs (c) Pristine NPs at H = 0.1 T (d) Pristine material with and without magnetic field for the last two peaks (plotted seperately because of low intensities)

Table 1. Peak position and their corresponding area and percentage under the fitted curves in Fig. 6.

| | Peak Position 1.84 eV | Peak Position 1.97 eV | Peak Position 2.12 eV | Peak Position 2.28 eV | Peak Position 2.44 eV | Total Area |
|---|---|---|---|---|---|---|
| AREA (H = 0 T) | 2264 | 5530 | 5881 | 4035 | 1789 | 19,499 |
| Percentage | 11% | 28.3% | 30.1% | 20.6% | 9.1% | 100% |
| AREA (H = 0.1 T) | 2026 | 4153 | 4903 | 3756 | 1649 | 16,487 |
| Percentage | 12.2% ↑ | 25.1% ↓ | 29.7% ↓ | 22.7% ↑ | 10% ↑ | 100% |
| AREA ($SnO_2$ + OTS) | 1031 | 3545 | 5143 | 4756 | 6849 | 21324 |
| Percentage | 4.8% ↓ | 16.6% ↓ | 24.1% ↓ | 22.3% ↑ | 32.1% ↑ | 100% |

Similar kind of changes, except for first peak, was observed for PL spectrum of pristine $SnO_2$ NPs taken in presence of an external magnetic field of H = 0.1 T (Fig. 5). In summary, application of external magnetic field and the surface functionalization of $SnO_2$ NPs had the same systematic effect of reduction in the emission at 1.97 and 2.12 eV with an increase at 2.28 and 2.44 eV channels (Table 1). The fractional area was simply proportional to the fractional number of electronic transitions in these five channels. Reasons for this observation is discussed below, which is intimately linked with the observed magnetic moments originating from defects in $SnO_2$ particles. It was reported in the time resolved X-ray induced PL study,[19] that the PL spectra with $E > 2.6$ eV had a very fast decay lifetime (below 10 nsec) indicating direct electronic transition from conduction band bottom to acceptor states. Whereas the decay time for PL emission below 2.6 eV was typically above 570 nsec indicating an initial trapped state of the electrons at some vacancies, slightly below the conduction band edge (shallow donor). Fig. S3, (supplementary information), shows the temperature dependence of PL for the pristine sample. The PL intensity in the energy window $E < 2.6$ eV increases progressively as the temperature decreases. This observation supports the transitions from the shallow donor levels.[21,22] There were several experimental studies showing shallow donor levels existing below the CBM.[19,23,24] Energy levels deep inside the band gap can appear from oxygen vacancies existing bulk of the NPs.[25,26] Here we present schematic band diagram to understand the existence of such large number of PL bands (Fig. 7). Surface oxygen vacancy related to shallow donor level denoted as $E_S$. There are 4 deep levels, with energies $E_1$, $E_2$, $E_3$ and $E_4$, which are defined as follows, $E_1 = E[O_V^B(1e^-)]$, $E_2 = E[O_V^B(2e^-)]$, $E_3 = E[O_V^P(1e^-)]$, and $E_4 = E[O_V^P(2e^-)]$. Where $E_1$ and $E_2$ are the levels for the first and the second electron at the in-plane oxygen (represented as $O^P$ in Fig. 2) site vacancy $O_V^P$ sites. Similarly, $E_3$ and $E_4$ are the levels for the first and the second electron at the bridging oxygen site (represented as $O^B$ in Fig. 2) vacancy $O_V^B$ sites. $E_2$ and $E_4$ are higher than $E_1$ and $E_3$, respectively because of necesesity for the extra coulomb repulsion energy to put the second electron (with opposite spin) at the vacancy site. Unlike the shallow donor level $E_S$, where the electron wave functions have finite extension over several lattice spacing, the deep level electron wave functions are far less extended. That is why the coulomb repulsion is important and the level for the first and second electron in such deep vacancy levels is considered to be different.[27] Various transitions with energy of the emitted photons are denoted in Fig. 7. Here, $E_2 - E_1 = 0.32$ eV, and $E_4 - E_3 = 0.31$ eV. These differences are roughly the coulomb repulsion energy which is discussed above. The transitions from $E_S \rightarrow E_1$ and $E_S \rightarrow E_3$ occur, when the bridging and the in-plane 'O' vacancy sites have no electrons. Hence electrons with any spin ↑ or ↓ from $E_S$ can come down with equal probability. The presence of an external magnetic field has no effect on these transition probabilities. The transitions from $E_S \rightarrow E_2$ and $E_S \rightarrow E_4$, occur on the other hand, when the bridging and planar O vacancy sites have already one electron. Under an external magnetic field, which polarizes the electron in the planar or bridging oxygen vacancy site along its direction (↑), then a ↑ spin electron from $E_S$ level cannot arrive there. In other words external magnetic field will suppress the number of transitions from $E_S \rightarrow E_2$ and $E_S \rightarrow E_4$, with emitted photon energies of 1.97 and 2.12 eV. Since the total number of emitted photons are conserved, there will be a corresponding rise in the emitted PL intensity in the, $E_S \rightarrow E_1$ and $E_S \rightarrow E_3$ transitions at energies 2.28 and 2.44 eV. Origin of the peak at 1.84 eV is not clear to us. It may be due to some kind of defect complexes related to hydroxyl or oxygenated species.[28] The intensity of the peak, however does not systemetically vary with fuctionlaization or with the applicaton of magnetic field (Table 1). This indicates that the defect at 1.84 eV does not take part in the spin polarization process. So we have not included the electronc transition related to it in the band diagram.

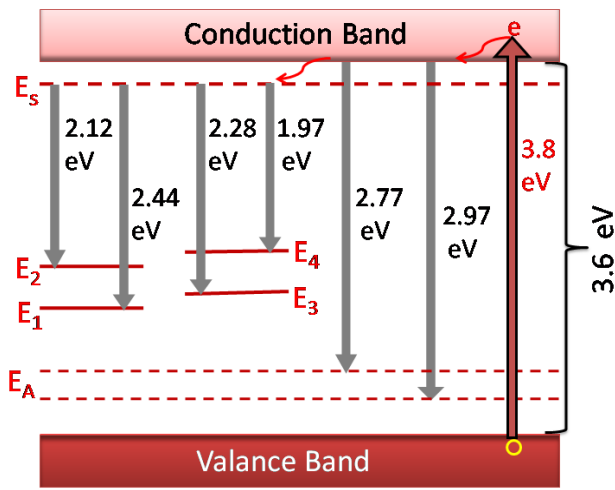

Fig. 7 Proposed band diagram of SnO$_2$ NPs showing different intra-band transition processes.

The PL intensity can be written as in the eqn (1),

$$I(\omega) \propto \sum_{i,j,\sigma} |<\Psi_{j,\sigma}(r_2)|e(r_1-r_2)|\Psi_{i,\sigma}(r_1)>|^2 \times \delta(E_{i,\sigma} - E_{j,\sigma} - \hbar\omega) \times f_{i,\sigma}(1-f_{j,\sigma}) \quad .....(1)$$

where $\Psi_{i,j,\sigma}$ is the wave function of the electrons at $i$ or $j$ sites, with energies $E_{i,j,\sigma}$ and the $f$'s are the corresponding Fermi functions. Magnetic field effect comes from both the dipole transition matrix element, and the occupation numbers (Fermi function) of the initial and final states. Out of all the electrons ($N_i$) at sites $i$, only $N_i e^{-t/\tau}$ reaches the sites $j$, thereby emitting a photon, where $\tau$ is the relaxation time for other non-radiative transition of the incoming electron. An external magnetic field essentially increases the flight path length of the electrons that reaches at $j$ sites, and thereby increases flight time $t$. Assuming a very little change in $\tau$ due to the application of magnetic field, this causes a decrease in the net number of radiative transitions, when a magnetic field is applied. Such flight time change under magnetic field was reported in case of porous silicon earlier.[29] This reduction of luminescence due to flight path deviation was same for all electronic transitions. Now let us look at the magnetic field dependence of emitted intensity arising from the modification of the occupation of initial and final electronic states. If a site $j$ has already two electrons (neutral vacancy) then no transition can occur to this state. If the site $j$ has no electrons (doubly charged vacancy), then electron of either spin from site $i$ can come there, and this transition does not have magnetic field dependence. But if the final state $j$ has one electron (singly charged vacancy), then there will be a magnetic field dependence for the radiative transition probability. Let $N_{i,j,\sigma}$ be the number of electrons at site $i, j$ with spin $\sigma$. In absence of magnetic field, $N_{i,j,\uparrow} = N_{i,j,\downarrow} = \frac{N_{i,j}}{2}$, with $N_{i,j} = N_{i,j,\uparrow} + N_{i,j,\downarrow}$. In presence of magnetic field, it can be written, $N_{i,\uparrow} = \frac{N_i}{2}+\delta_1$, $N_{i,\downarrow} = \frac{N_i}{2}-\delta_1$, and similar expression for $j$ sites with $\delta_1$ replaced by $\delta_2$, where $\delta_{1,2}$ is proportional to external magnetic field. The net number of radiative transitions in this channel is proportional to, $(\frac{N_i}{2}+\delta_1) \times (\frac{N_j}{2}-\delta_2) + (\frac{N_i}{2}-\delta_1) \times (\frac{N_j}{2}+\delta_2) = \frac{N_i N_j}{2} - 2\delta_1\delta_2$. A suppression of radiative transition in the channels will result where the final state at site $j$ has one electron already. In short, an overall decrease in the number of radiative transitions, in all channels due to increase in flight path (leading to decay in non-radiative processes) is now expected in presence of magnetic field. Amongst the surviving radiative transitions, the number of transitions where the final state has already one electron will decrease in the presence of magnetic field. The number of radiative transition where the final state has no electron present will increase in presence of magnetic field, as the total number of surviving radiative transitions is conserved.

This is exactly what was observed in present experiments. Process of surface functionalization creates extra moments on the surface of the particles (Fig. 2). When the planar or bridging oxygen vacancy sites have one electron (with a net spin 1/2) these vacancy electron moments will have exchange interaction with the surface moment created by functionalization at nearby oxygen sites. This was discussed previously while describing anomalous magnetism in surface functionalized SnO$_2$ NPs in the present investigation (Fig. 3). So there will be an effective internal field (coming from the exchange interaction) acting on the planar/bridging vacancy site electron. Situation is exactly the same, as if an external field is applied. This explains why both surface functionalization and external magnetic field has almost same effect on the PL transitions (Fig. 5). We assign peaks at 2.77 eV and 2.97 eV to the transitions from conduction band bottom to the acceptor states, which reside in the region of 0.55-0.7 eV above the valence band top, as also supported by literature.[19,27]. An interesting finding in the present studies is that, these emitted lines are very faint in pristine $SnO_2$ NPs, but as the $SnO_2$ NPs are functionalized with OTS, the intensity in these channels increases substantially (Fig. 5). This fact is correlated with an overall reduction in all transitions in the energy window E < 2.6 eV that was discussed earlier. Some Si atoms bond with surface oxygen vacancies (shallow donors) leading to a loss of shallow donor concentrations. This loss in addition to surface passivation of dangling bonds of $SnO_2$ surfaces might be the reason for an overall decrease in PL intensity of all lines at E < 2.6 eV. Since the number of electronic transitions are conserved, this will mean that there will be a corresponding increase in the transitions that starts from the conduction band bottom, i.e. in the energy window E > 2.6 eV. This may be the dominant reason for the considerable increase in PL intensity for E > 2.6 eV after functionalization of the particles. To summarize, we have seen that surface functionalization of $SnO_2$ NPs by OTS creates unpaired electrons on the surface carrying net magnetic moment (Fig. 3 and discussion therein). From PL spectra, there are four different kinds of defect related energy levels within the gap (1) shallow donor levels (due to surface oxygen vacancy), (2) and (3) deep levels due to oxygen vacancies at bridging and planar sites, and (4) acceptor levels due to metal atom vacancies. Both external magnetic field and surface functionalization decrease the number of transitions with energies 1.97 and 2.12 eV, and increases transitions with energies 2.28 and 2.44 eV. All these phenomena, are intimately related to the type of chemical bonding on the surface, which creates magnetic moment on the surface of functionalized material.

## Conclusions

In conclusion, a high magnetic moment was observed in the pristine $SnO_2$ NPs at room temperature. An anomalous enhancement in magnetic moment and PL intensity were achieved by surface functionalization with silane (OTS). Changes in the electronic configuration in introducing large magnetic moment by the functionalization process is demonstrated using magnetic field dependent PL studies. Thus, this study demonstrate that a complete understanding of electronic processes involved in controlling optical properties and in the evolution of the magnetic moment using suitable functionalization of $SnO_2$, can lead to the development of promising oxide based spintronic materials and optoelctronic applications.

## Acknowledgements

We thank Dr. John Philip, SMARTS, MMG, IGCAR for the support in magnetic measurements and Dr. S. Amirthapandian, MPD, MSG, IGCAR for recording TEM images.

## Notes and references

[a] Surface and Nanoscience Division,
Indira Gandhi Centre for Atomic Research, Kalpakkam-603102, India.
[*]E-mail : dasa@igcar.gov.in,

[b] Materials Physics Division,
Indira Gandhi Centre for Atomic Research, Kalpakkam-603102, India.
[*]E-mail : manas@igcar.gov.in,

Electronic Supplementary Information (ESI) available: [Raman spectra of functionalized NPs and temperature dependent PL. This material is available free of charge via the Internet at]. See DOI: 10.1039/b000000x/

# Supporting information

Octadecyltrichlorosilane (OTS) molecules are linked to $SnO_2$ nanoparticles (NPs) through a self-assembly process by forming covalent bonds. Figure S1 shows the Raman spectra of OTS treated $SnO_2$ NPs, which show the allowed modes for rutile $SnO_2$ phase along with molecular vibrations from OTS molecules. Peak around 570 $cm^{-1}$(*D*) is due to in-plane 'O' vacancy.[1] Raman spectroscopy has been widely used to understand ordered and disordered phases of the covalently bonded OTS to silica surfaces.[2,3] Molecular vibrations originating from the breadth and asymmetry of $CH_2$ twist ($\tau$-$CH_2$) and bending of $CH_2$ ($\delta$-$CH_2$) modes were observed at 1293, and 1441 $cm^{-1}$, respectively (Fig. S1). Similarly symmetric ($\upsilon_s$) and asymmetric ($\upsilon_a$) stretching modes of $CH_2$ and $CH_3$ modes were also shown in Figure S1. Vibrational modes of $\upsilon_s(CH_2)$ and $\upsilon_a(CH_2)$ were observed at 2846 and 2881 $cm^{-1}$, respectively whereas the same for $\upsilon_s(CH_3)$ and $\upsilon_a(CH_3)$ appeared at 2935 and 2962 $cm^{-1}$, respectively. In the ordered phase of OTS, $\tau$-$CH_2$ mode was reported to shift to lower frequency in comparison to the disordered phase of OTS at 1304 $cm^{-1}$. The $\tau$-$CH_2$ mode at 1294 $cm^{-1}$ supports the formation of an ordered OTS on the $SnO_2$ surfaces. Moreover, the intensity ratio of $\upsilon_a(CH_2)$ to $\upsilon_s(CH_2)$ is described to be higher than 0.9 for the ordered phase. In the present study, intensity ratio of the above modes exhibits around 1.3 extending support to the formation of ordered OTS on $SnO_2$ surfaces. The intensity ratio, however is found to be lower than the reported value of 1.6 - 2 for the crystalline phase of OTS on micron sized silica surfaces.[2,3] In the present context, spherical NPs of $SnO_2$ allows only limited surface area for the ordering to be observed.

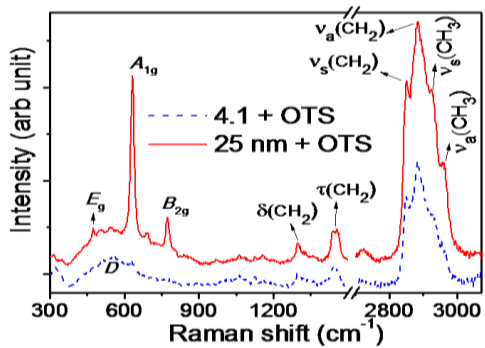

Fig. S1 Raman spectra of OTS treated $SnO_2$ nanoparticles of 4 nm, and 25 nm

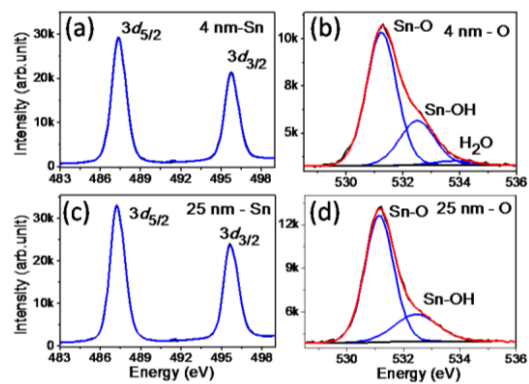

Fig. S2 XPS spectra of (a) Sn 3d (b) O 1s of $SnO_2$ (4 nm) and (c) Sn 3d (b2) O 1s of $SnO_2$ (25 nm).

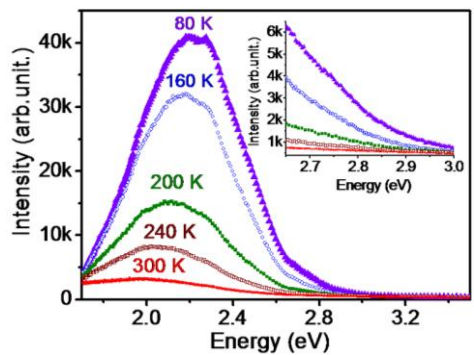

Fig. S3 Temperature dependent photoluminescence of pristine 25 nm $SnO_2$ nanoparticles. Inset showing the zoomed region between 2.65 to3 eV.

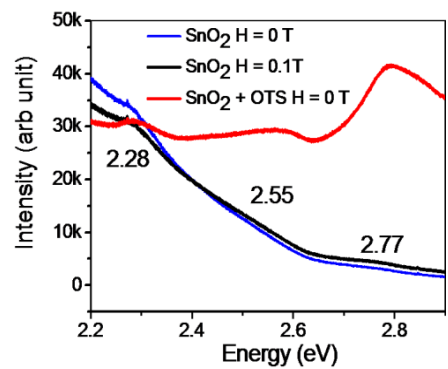

Fig. S4 Zoomed region of Fig. 4 showing no new peak after functionalization.